\documentclass[twocolumn,showpacs,preprintnumbers,amsmath,amssymb]{revtex4}
\usepackage{graphicx}
\usepackage{dcolumn}
\usepackage{bm}
\begin{document}

\title{{Coulomb scattering in plasma revised}}

\author{S.~Gordienko}
\affiliation{L.D.Landau Institute for Theoretical Physics, Russian Academy of Science,
Kosigin St. 2, Moscow, Russia}
\email{gord@itp.ac.ru}
\author{D.V.~Fisher}
\affiliation{Faculty of Physics, Weizmann Institute of Science - Revohot 76100, Israel}
\email{fndima@plasma-gate.weizmann.ac.il}
\author{J.~Meyer-ter-Vehn}
\affiliation{Max-Planck-Institut f\"ur Quantenoptik - D-85748 Garching, Germany}
\email{meyer-ter-vehn@mpq.mpg.de}

\date{\today}

\begin{abstract}
{A closed expression for the momentum evolution of a test particle
in weakly-coupled plasma is derived, starting from quantum many
particle theory. The particle scatters from charge fluctuations
in the plasma rather than in a sequence of independent binary
collisions. Contrary to general belief, Bohr's (rather than
Bethe's) Coulomb logarithm is the relevant one in most plasma
applications. A power-law tail in the distribution function is
confirmed by molecular dynamics simulation.} 
\end{abstract}


\maketitle

Though Coulomb scattering is a most basic process in plasma and
has been studied for a century \cite{Bohr}, doubts concerning the
treatment as a sequence of independent binary collisions remained
\cite{Kog,Siv}, and recent analysis \cite{Gor} has revealed  that
this standard assumption is not justified in general and requires
revision. Here we derive the time-dependent many-particle
wavefunction of a test particle simultaneously interacting with
$N$ particles residing in the Debye sphere. The plasma parameter
$N=n\lambda^3>1$ involves the plasma density $n$ and the screening
length $\lambda=max(v_T/\omega_p,v_0/\omega_p)$, where $v_T$ is
the thermal velocity of electrons, $v_0$ the velocity of the test
particle, and $\omega_p=\sqrt{4\pi n e^2/m_e}$ the plasma
frequency. We emphasize that the collective interaction described
here is different and in addition to Debye screening; it is not
included in the usual dielectric approach \cite{Larkin}. The new
results can be viewed as interaction of the test particle with the
charge fluctuations inside the Debye sphere; in this picture, the
test particle of charge $Z_0e$ is scattered by an effective,
spatially extended charge $e\sqrt{N}$ rather than by a sequence of
binary collisions.

\begin{figure}
{\scalebox{0.5}{\includegraphics{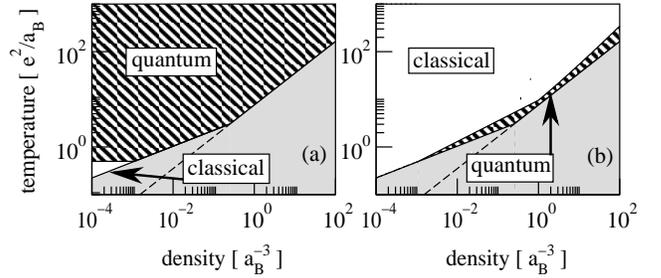}}}
\caption{ Regions in a density-temperature plane
(atomic units) in which Bohr's classical Coulomb logarithm (white
area) and Bethe's quantum expression (hatched area) apply; (a)
binary collision theory with borderline defined by
$\alpha=Z_0e^2/\hbar v_0=1$, and (b) present theory with approximate
separation along $\alpha^2N(1+\ln N)=1$. Also shown as grey area
are the region of strongly non-ideal plasma (borderline: $T\sim
n^{1/3}$) and the region of degenerate plasma (borderline: $T\sim
n^{2/3}$).} 
\label{fig1}
\end{figure}

This has deep consequences for the Coulomb logarithm, because it
drastically shifts the borderline between classical and quantum
Coulomb scattering, extending the domain in which the classical
approximation applies. This is shown in Fig. \ref{fig1}. In the binary
collision approach (Fig. \ref{fig1}a), the borderline is given by the
parameter $\alpha=Z_0e^2/\hbar v_0$ such that $\alpha<1$ defines
the quantum-mechanical region where the Born approximation leads
to Bethe's logarithm $L_q=\ln(\lambda m_e v_0/\hbar)$
\cite{Bethe}, while for $\alpha>1$ classical mechanics apply
leading to Bohr's logarithm $L_{cl}=\ln(\lambda m_ev_0^2/Z_0e^2)$.
In the present theory instead, the borderline is given by $\alpha
N^{1/2}\approx 1$, and this leads to a very different picture in
Fig.1b. Now $L_{cl}$ applies to almost the entire high-temperature
region, including e.g. the important domain of magnetic fusion
plasmas, while $L_q$ plays only a marginal role.

Let us first discuss this result in qualitative terms. A
particular feature of the Coulomb ($1/r$) interaction is of
crucial importance in this case, namely that scattering does not
depend on $\alpha$ and $\hbar$, as we know from the Rutherford
cross section. The difference between $\alpha<1$ and $\alpha>1$
regions arises only when the potential deviates from $1/r$, as it
is the case in a plasma due to Debye screening occurring at long
distances $\lambda$. This means the distinction between classical
and quantum treatment reveals itself for small-angle scattering,
while close collisions with large-angle scatter are not affected.
This point has been emphasized strongly by Bohr (see p.448 in \cite{Bohr})
and also in Sivukhin's review \cite{Siv} (p. 109--113).
In accordance with Bohr ``any attempt to attribute
the difference between [the classical $\alpha\gg1$
and quantum $\alpha\ll1$ cases] to the obvious failure of [the classical]
pictures in accounting for collisions with an impact parameter smaller
than [the de Broglie wave--length] will be entirely irrelevant. In fact,
this argument would imply a difference between two distribution
for the large angle scattering, while the actual differences
occur only in the limits of small angles.''

Now let us compare the classical scattering angle
$\delta_{cl}=(Z_0e\, e_{*}/m_0v_0^2)/\lambda$ at distance
$\lambda$ with that of quantum diffraction
$\delta_q=(\hbar/m_0v_0)/\lambda$ \cite{Lan}. The open question
here concerns the effective net charge $e_{*}$ which the test
particle experiences when passing the Debye sphere. The value of
$e_{*}$ is not evident, because we are dealing with Coulomb
collisions at distances much larger than $1/n^{1/3}$. The binary
collision approximation circumvents this predicament by alleging
that the total scattering can be treated as the sum of independent
binary interactions happening at different times \cite{Siv}. One
is then led to take $e_{*}=e$ for granted instead of actually
calculating $e_{*}$. The central result of this paper will be that
the effective charge is essentially given by $e_{*}\approx
eN^{1/2}$. The matching condition then is $\delta_{cl}/\delta_q
\approx \alpha N^{1/2} \approx 1$ replacing the condition
$\alpha\approx 1$ obtained in binary collision approximation
\cite{Siv}. The theory underlying Fig.\,1b will now be derived. As
a central result, we also present molecular dynamics simulations
confirming the analytic theory.

The present analysis starts from a full quantum-mechanical
description of the plasma in terms of the many-particle
wave-function $\psi=\exp(iS /\hbar)$. The action function $S$
satisfies the equation
\begin{equation}
\label{eq:1}
-\frac{\partial S}{\partial t}=
\sum_j\left(\frac{(\nabla_{j}S)^2}{2m_j}+\sum_{k>j}
U_{j,k}-i\hbar \frac{\Delta_{j}S}{2m_j}\right),
\end{equation}
where the indices $j$ and $k$ denote plasma particles for
$j,k=1,2, 3\dots$ and the test particle for $j,k=0$, $m_j$ are the
masses, and $U_{j,k}$ represent the Coulomb interactions. Eq. (\ref{eq:1})
is equivalent to the exact Schr\"odinger equation and has the form
of a Hamilton-Jacoby equation with additional terms that are
proportional to $\hbar$ and describe quantum effects. We examine
the solution of Eq. (\ref{eq:1}) for the particular initial conditions
$S(t=0)=\sum_{j\ge0}{\bf p}_j\cdot{\bf r}_j$,
(the Green function for the coordinate--momentum representation).
where ${\bf p}_j$ are given momenta of the plasma particles at $t=0$,
and introduce
\begin{equation*}
\sigma = S - \sum_{j\ge0}\left({\bf p}_j\cdot{\bf r}_j-{\bf p}_j^2t/2m_j\right)
\end{equation*}
such that $\sigma(t=0)=0$.
Now the clue for solving Eq.(\ref{eq:1}) is the high-energy approximation
\cite{Lan} which applies to an almost ideal plasma and requires
$|{\bf p}_j|\gg|\nabla_j\sigma|$. Under this approximation
we find
\begin{equation}
\label{eq:2}
-\frac{\partial \sigma}{\partial t}=
\sum_j\left({\bf v}_j\cdot\nabla_{j}\sigma
+\sum_{k>j} U_{j,k}-i\hbar
\frac{\Delta_{j}\sigma}{2m_j}\right),
\end{equation}
where ${\bf v}_j={\bf p}_j/m_j$.
The solution of Eq. (\ref{eq:2}) for $\sigma(t=0)=0$ is
\begin{equation}
\label{sol}
\sigma=-\sum_{j\ge0}\sum_{k>j} \int_0^t U_{k,j}\left({\bf D}_{k,j}\right)
\,d\tau,
\end{equation}
where $\delta {\bf r}_{k,j} ={\bf r}_k-{\bf r}_j$, $\delta {\bf
v}_{k,j} ={\bf v}_k-{\bf v}_j$ and ${\bf D}_{k,j}=\delta{\bf
r}_{k,j}-\delta{\bf v}_{k,j}(t-\tau)$. It can be verified by
direct insertion. The terms $\Delta_{j}\sigma$ proportional to
$\hbar$ vanish for the special case of Coulomb $U\propto 1/|{\bf
D}|$ interactions for distances $|{\bf D}|>0$.

The problem of solution (\ref{sol}) is that it contains the singularities of
$U_{k,j}({\bf D}_{k,j})$ for close encounters with
$|{\bf D}_{k,j}|\rightarrow 0$.
This deficiency is due to the high energy approximation.
Inserting Eq. (\ref{sol}) into $|{\bf p}_j|\gg|\nabla_j\sigma|$,
we find that this condition is fulfilled only for regions
\begin{equation}
\label{shortcutoff}
|{\bf D}_{k,j}|>|e_je_k| /(\mu_{k,j} |\delta {\bf v}_{k,j}|^2)
\end{equation}
with $0<\tau<t$ and $\mu_{k,j}=m_jm_k/(m_j+m_k)$. Had we solved
the nonlinear equation (\ref{eq:1}) exactly, we had obtained a
non--singular result with the maximum momentum transfer of
$2\delta v_{j,k}\mu_{j,k}$, as we know from Rutherford scattering.
The way to deal with this problem is to cut out in the
wavefunction those spatial regions which do not satisfy Eq. 
(\ref{shortcutoff}).
The cut-off (\ref{shortcutoff}) warrants that the maximum momentum transfer
$2\delta v_{j,k}\mu_{j,k}$ is preserved; this can be verified by
operating with $-i\hbar\nabla_j$ on $\exp(iS/\hbar)$. It should be
understood that this short-range cut-off is a technical
correction (compare with \cite{Bohr}, p. 448--449:''\dots the central region of the field\dots, which, on classical mechanics, is responsible for all large angle scattering will, for $\alpha\ll1$,\dots gives rise to only a fraction of the order $\alpha^4$ of the Rutherford scattering''). It has nothing in common with differences between
classical and quantum scattering. These reveal themselves only at long ranges
\cite{Bohr,Siv}. Another detail concerning the general
wavefunction concerns the initial conditions. In case the initial
state of the plasma is defined by the wavefunction $\phi(t=0,{\bf
p}_1,{\bf p}_2,\dots)$
 rather than by a fixed set of momenta,
the corresponding general wavefunction is given by 
\begin{eqnarray*}
\psi(t,{\bf
r}_0,{\bf r}_1,\dots)= \int\exp(iS/\hbar) \phi(t=0,{\bf p}_1,{\bf
p}_2,\dots)\prod_{k\ge1}\,d{\bf p}_k.
\end{eqnarray*} 
As it turns out, the
explicit form of $\phi$ is of no relevance in the applications
discussed below.

We now have at our disposal in analytical form
the time-dependent many-particle wavefunction
describing a dilute high-temperature plasma. This is a remarkable result.
An outstanding feature is that it describes simultaneous
multiple Coulomb interactions between the particles and, in this respect,
goes beyond the binary collision approximation.
Another essential property is that it holds
for both the quasi-classical regime ($\sigma\gg\hbar$) as well as
the deeply quantum-mechanical regime ($\sigma\ll\hbar$)
and therefore provides a unique tool to investigate the
transitional region.
We now proceed to use this wavefunction to calculate
plasma properties. This is straightforward, though tedious,
and therefore we can give here only the main results, leaving
technical derivations to a separate publication.

We first consider the distribution function ${ M}(t,{\bf Q})$
of transverse momentum ${\bf Q}$ of a test particle moving at time $t=0$
with momentum ${\bf p}$ collinear to the $x$-axis.
For brevity we  consider
a fast ion $m_0\gg m_e$ , $v_0\gg v_T$ for times
that are longer than $t_0=\lambda/v_0$, though shorter than the
collision time $t_c\sim m_0Nt_0/m_eL_{cl,q}$. ${ M}(t,{\bf Q})$
is obtained as the matrix element
\begin{equation}
\label{M}
M(t,{\bf Q})=\int\exp\left(i{\bf Q}\cdot{\bf R}/\hbar\right)
F(t,{\bf R})\,d^2{\bf R}/(2\pi\hbar)^2
\end{equation}
where
\begin{eqnarray*}
F(t,{\bf R})=
\frac{1}{V}\int\psi(t,{\bf r}_0,{\bf r}_1,\dots) \psi^*(t,{\bf
r}_0+{\bf R},{\bf r}_1,\dots) \prod_{k\ge0}\,d^3{\bf r}_k
\end{eqnarray*}
and $V$ is the plasma volume.

Expression (\ref{M}) can be significantly simplified for the case under
consideration. The test particle affects only plasma particles in
the interaction sphere $|\delta{\bf r}_{0,j}|<\lambda$, for which
two--particle correlations among plasma particles are small owing
to $T\gg e^2n^{1/3}$. Aiming deliberately for calculations with
logarithmic accuracy, we can omit the integration over
$|\delta{\bf r}_{0,j}|>\lambda$ and use the method developed in
\cite{Gor}. We find $F(t,{\bf R})=F_e(t,{\bf R})F_i(t,{\bf R})$
where
$F_{e}(t,{\bf R})=\exp(-f_{e}(t,{\bf R}))$,
\begin{eqnarray}
\label{f_e}
f_e=2N\int\sin^2\left[\frac{\alpha}{4}\left(g(t,{\bf r}_0+{\bf R})-
g(t,{\bf
r}_0\right)\right]\frac{d^3{\bf r}_0}{\lambda^3},\\
\label{g}
g=\int_{-v_0t}^{v_0t}V_0(x_0+\zeta/2-v_0t,y_0+Y,z_0+Z)\,d\zeta,\
\end{eqnarray}
$N=n\lambda^3$, $\alpha=Z_0e^2/\hbar v_0$, ${\bf r}_0=
(x_0,y_0,z_0)$, $V_0({\bf r}_0)=1/|{\bf r}_0|$ for $|{\bf r}_0|
<\lambda$ and $V_0=0$ for $|{\bf r}_0|>\lambda$. In the ${\bf
r}_0$--integration, the domain
$\min((y_0+Y)^2+(z_0+Z)^2,y_0^2+z_0^2)<r^2_{cl}=(Z_0e^2/m_ev_0^2)^2$
is excluded for reasons discussed above. The ion function $F_i$
has the same structure as $F_e$ and is simply obtained by
substituting ion parameters. ${ M}$ is the convolution of ${ M}_e$
and ${ M}_i$, where 
\begin{eqnarray*}
{ M}_{e,i}(t,{\bf Q})=\int\exp(i{\bf
Q}\cdot{\bf R}/\hbar)F_{e,i}(t,{\bf R})\,d^2{\bf R}/(2\pi\hbar)^2
\end{eqnarray*}
are the transverse momentum distributions due to the
electron--projectile and ion--projectile interaction. In the
following, most of the discussion is restricted to $M_e$.

Let us discuss the structure of Eqs.\,(\ref{f_e}),(\ref{g}),
which are presented here for the first time. The detailed analysis
is quite intricate, and here we give only the main results without
derivation. We observe that only small enough $f_e$ can contribute
to $M$ and that therefore, owing to the large multiplier $N$ in
Eq.\,(\ref{f_e}), the sin$^2$-term needs to be small and can be
expanded. Then $M$ depends essentially on the parameter
combination $\alpha^2N$ only; more rigorous analysis gives
$\gamma=\alpha^2N\ln N$. The quantum regime is restricted to
$\gamma <1$, while the classical regime is found for $\gamma > 1$
and will be discussed first. Evaluating Eq.\,(\ref{f_e}) in the
limit of very small $R$, one finds $f_e({\bf R})=\nu t \ln(\lambda
m_e v_0^2/Z_0e^2)({\bf R}/\hbar)^2$ with $\nu={2\pi n Z_0^2
e^4}/v_0$. This is the relevant region in the Fourier integral of
${M}_e$ for large enough time $t$ ($t_1\ll t<t_c$, see below). The
function $F_e({\bf R})$ is then a Gaussian, and ${M}_e$ can be
easily calculated. Setting $F_i=1$, we obtain $\left<Q^2\right>_e
=\int {\bf Q}^2M_e(t,{\bf Q})\,d^2{\bf Q}= 4\nu t L_{cl}$ and
recover the classical Coulomb logarithm $L_{cl}=\ln(\lambda m_e
v_0^2/Z_0e^2)$. The important new result here is that it applies
to the whole region $\gamma > 1$ and not just to $\alpha>1$.

It should be noticed, however, that the function $f_e({\bf R})=\nu
t \ln(\lambda/{\tilde R})({\bf R}/\hbar)^2$ is more complicated in
general and contains a  factor $\ln{\tilde R}$ for larger radii,
where ${\tilde R}=\max(\alpha|{\bf  R}|,\alpha\hbar/m_e v_0)$ for
$\alpha>1$ and ${\tilde R}=\max(|{\bf R}|,\alpha\hbar/m_e v_0)$
for $\alpha<1$. For short times, just somewhat larger than
$\lambda/v_0$, this logarithmic factor modifies the Gaussian
character of $M_e(t,{\bf Q})$, giving it a power--law tail at high
$Q=|{\bf Q}|$. We then obtain
\begin{equation}
M(t,{\bf Q})=\exp(-Q^2/2p_0^2) /(2\pi p_0^2)
\end{equation}
for $Q^2 < 2p_0^2 \ln\Lambda$ and
\begin{equation}
M(t,{\bf Q})=2p_0^2/(\pi \Lambda Q^4)
\end{equation}
for $2p_0^2\ln\Lambda< Q^2 < (2m_e v_0)^2$, where $\Lambda$ is a
solution of $\Lambda=\ln(2\pi\alpha_1 n \Lambda \lambda^2 v_0
t)\gg1$, $\alpha_1=\min(1,\alpha)$ and $p_0^2=\nu \Lambda t$.
Physically, the Gaussian distribution at small $Q$ corresponds to
small angle scattering and the power--law tail to close collisions
with large momentum transfer. The tail is obtained only as long as
$t<t_1=2m_e^2v_0^2/\nu \Lambda\ln\Lambda$, such that
$2p_0^2\ln\Lambda <(2m_e v_0)^2$. For longer times, small angle
scattering dominates both small and large $Q$ regions and the tail
disappears.

\begin{figure}
{\scalebox{0.35}{\includegraphics{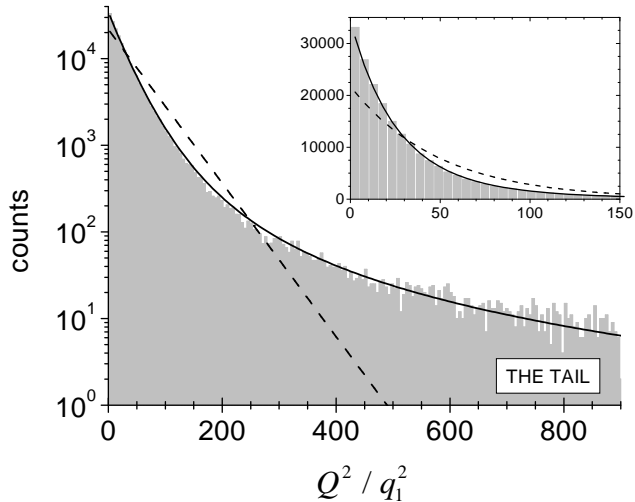}}}
\caption{ Comparison between
MD simulation (histogram), present theory (solid line), and
predictions of the traditional diffusion approximation (dashed
line); $\pi M(t,Q)$ is plotted versus $Q^2/q_1^2$ for time
$t=2l/v_0$, $l=7.239\times 10^{-6}$cm, velocity of test particle
$v_0=2.297\times10^{8}$cm/s, and $q_1^2=2 \pi n Z_{0}^{2}e^{4}
t/v_0$. The insert shows the same plot, but with linear scale and
zoomed to low $Q$. The simulation assumes an equal number of
randomly distributed, fixed Coulomb centers of opposite charge
$\pm e$ and densities $n_{+}=n_{-}=n=1.054\times 10^{17}$
cm$^{-3}$; the cold plasma limit is chosen with thermal velocity
$v_{th}\ll v_0$. The plasma volume is taken as $V=6l\times
2l\times 2l$ with the test particle ($Z_0=2$, $m_0=m_e$) moving
along the central axis in $x$-direction and starting at a distance
$2l$ from the surface. The trajectory of the test particle,
interacting with all Coulomb centers, is obtained by solving the
classical equation of motion by a second-order scheme with an
adaptive time step. The histogram corresponds to $2.06\times 10^5$
independent trials. The solid line has been obtained numerically
from  Eqs.\,(\ref{f_e})--(\ref{g}) with $F_i=F_e=\exp(-f_e)$. The
screening length is set to $\lambda=l$ such that the finite plasma
volume seen by the test particle in this model simulation just
mimics the physically screened volume occuring in reality. The
straight dashed line is the Landau collision integral prediction
$M(t,Q)=\exp(-Q^2/Q_0^2)/ (\pi Q_0^2)$ with $Q_0^2=8\pi
e^4Z_0^2(n_++n_-)L_{0}t/v_0$; $L_{0}=6.5$ is the classical Coulomb
logarithm evaluated for the parameters of the simulation. }
\label{fig2}
\end{figure}

We have checked the occurrence of this power-law tail by molecular
dynamics (MD) simulations. We consider a test particle with
$v_0\gg v_{th}$ scattered completely classically in a finite
plasma volume, having dimensions $l$ of the order of the screening
length. The simulation has been performed for a model case with
$N=80$ and $\alpha=1.9$, just feasible on a modern PC. Results are
plotted in Fig.\,\ref{fig2} for time $t=2l/v_0<t_1$ . The histogram
presenting the MD results is in best agreement with the present
theory (solid curve), clearly showing the power-law tail at high
momenta. For comparison, also the purely Gaussian distribution
obtained from the Landau collision integral is given as dashed
line. Details of these simulations are outlined in the caption.
Here we should make it clear that the power law tail originating
from close collisions is obtained in nearly identical form within
the binary collision approach, as it was shown by Landau
\cite{Landau} and Vavilov \cite{Vavilov}. The present theory
differs for small-angle scattering and therefore in the Gaussian
part of $M(t,Q)$. To show the difference quantitatively, we have
also solved the kinetic equation used in \cite{Landau,Vavilov}.
The result can be written in a form equivalent to Eq. (\ref{M})
with the function $f_e$ now given approximately by $f_e^{(b)}=\nu
t \ln(\alpha_1\lambda/{\tilde R})({\bf R}/\hbar)^2$, where
${\tilde R}=\max(\alpha|{\bf R}|,\alpha\hbar/m_e v_0)$ and
$\alpha_1=\min(1,\alpha)$. The effect of the present theory is
that the Gaussian part grows more rapidly. This is consistent with
enhanced small-angle scattering due to simultaneous interaction
with many plasma particles.

We have seen in Fig.\,\ref{fig1} that the quantum limit ($\gamma<1$) is
relevant only in a marginal parameter region. Nevertheless, it is
contained in $M_e$. For $\gamma<1$, one can use first-order
expansions of $F_e=1-f_e$ and of the sin$^2$-term in $f_e$ to
find, after some algebra, $M_e(t,{\bf Q})=C(t)\delta({\bf
Q})+\sigma(Q)nvt/(m_ev)^2$, where
$C(t)=1-nvt/(m_ev_0)^2\int\sigma(Q)\,d^2{\bf Q}$ and
$\sigma(Q)\approx r_{cl}^2(2m_e v_0)^4/(Q^2+(\hbar/\lambda)^2)^2$
is the cross--section of the screened Coulomb potential. This
leads to  $\left<Q^2\right>_e = 4\nu t L_{q}$ and the quantum
(Bethe) logarithm $L_q=\ln(\lambda m_ev_0/\hbar)$. This
first-order Born result is obtained here for $\alpha N^{1/2}<<1$,
but not for $\alpha <1$ in general. This may be understood
qualitatively looking at second-order processes. Consider the
perturbation of the interaction of the test particle with a plasma
particle $j$ by another plasma particle $k$. This effect is small
of order $\alpha^2$, but since for a plasma with long--range
Coulomb forces $N$ particles contribute to this second-order
process, it can be neglected only if $\alpha^2N<1$.

Let us finally calculate the energy ${\mathcal
E}(t)=\left<\psi\left|{\hat H}_p\right|\psi\right>$ the plasma
gains due to the energy loss of the test particle. Here 
\begin{equation*}
{\hat
H}_p=\sum_{j\ge1}\left(
-\hbar^2\Delta_j/(2m_j)+\sum_{k>j}U_{j,k}\right)
\end{equation*}
 is the
Hamiltonian of the plasma without the test particle \cite{Larkin}
and $\psi(t,{\bf r_0},{\bf r}_1,\dots)$ the full wavefunction.
Making use of the same approximations as in the derivation of Eq.
(\ref{M}), straightforward algebra leads to $d{\mathcal
E}/dt=\left<Q^2\right>_e/2m_e + \left<Q^2\right>_i/2m_i$, where
the first term is the contribution from plasma electrons with mass
$m_e$ and the second  from ions with mass $m_i$. The corresponding
electron part of the stopping power is then found in the standard
form 
\begin{eqnarray*}
\frac{dE}{d x}= -\frac{1}{v_0}\frac{d{\mathcal E}}{d t}=-\frac{4\pi n e^4 Z_0^2}{m_e
v_0^2}L,
\end{eqnarray*} 
but now with $L=L_{cl}$ for $\gamma>1$ and $L=L_q$ for $\gamma<1$.

In conclusion, it has been shown that the theory of Coulomb
scattering in dilute plasma needs revision. Bohr's classical
Coulomb logarithm $L_{cl}$ is found to apply for
$\alpha\sqrt{N}>1$ rather than $\alpha>1$, and this covers most of
the density-temperature plane, relevant to practical applications.
This result calls for experimental verification. We propose to
measure energy loss of fully stripped ions in carefully
characterized, fully ionized plasma layers. The parametrically
different dependence of $L_{cl}$ and $L_q$ on ion charge $Z_0$ and
velocity $v_0$ should allow for a clear distinction.

\acknowledgements
The authors acknowledge controversial discussions with M.~Basko
and G.~Maynard. This work was supported by Bundesministerium f\"ur
Forschung und Technologie, Bonn and Counsel for the Support of Leading Russian Scientific Schools 
(Grant  No. SS-2045.2003.2).

\end{document}